\documentclass[prd,showpacs]{revtex4}
\usepackage{graphicx}
\usepackage{amssymb}
\begin{document}
\title{Operating LISA as a Sagnac interferometer}
\author{Daniel A.~Shaddock}
\address{Jet Propulsion Laboratory, California Institute of Technology, Pasadena, 91109.}
\date{\today}

\begin{abstract}
A phase-locking configuration for LISA is proposed that provides a significantly simpler mode of operation.  The scheme provides one Sagnac signal readout inherently insensitive to laser frequency noise and optical bench motion for a non-rotating LISA array. This Sagnac output is also insensitive to clock noise, requires no time shifting of data, nor absolute arm length knowledge.  As all measurements are made at one spacecraft, neither clock synchronization nor exchange of phase information between spacecraft is required. The phase-locking configuration provides these advantages for only one Sagnac variable yet retains compatibility with the baseline approach for obtaining the other TDI variables. The orbital motion of the LISA constellation is shown to produce a 14~km path length difference between the counter-propagating beams in the Sagnac interferometer. With this length difference a laser frequency noise spectral density of 1 Hz/$\sqrt{\rm Hz}$ would consume the entire optical path noise budget of the Sagnac variables. A significant improvement of laser frequency stability (currently at 30~Hz/$\sqrt{\rm Hz}$) would be needed for full-sensitivity LISA operation in the Sagnac mode. Alternatively, an additional level of time-delay processing could be applied to remove the laser frequency noise. The new time-delayed combinations of the phase measurements are presented.
\end{abstract}

\maketitle

\section{Introduction}
The Laser Interferometer Space Antenna (LISA) \cite{LISA} is a joint NASA-ESA project to detect and study low frequency gravitational waves. LISA consists of three spacecraft separated by 5 million kilometers flying a total of six proof masses in heliocentric drag free orbits. The change in separation of each proof mass pair must be measured to a level of approximately 20~pm/$\sqrt{\rm Hz}$. One of the main challenges of the LISA interferometry is ensuring that the laser frequency noise ($\Delta \nu/\nu\sim 10^{-13} /\sqrt{\rm Hz}$) does not obscure the gravitational wave induced length change ($\Delta L/L\sim10^{-20}/\sqrt{\rm Hz}$). Time-delay interferometry (TDI) \cite{TintoPRD99,ArmstrongAPJ99} is a time domain post-processing technique that provides several interferometer outputs that are free from laser frequency noise. Two categories of TDI combinations have been examined: the Sagnac variables, and the Michelson variables. The Sagnac variables $\alpha$, $\beta$, $\gamma$, and $\zeta$, are generators for the entire space of frequency noise-free interferometric combinations \cite{ArmstrongAPJ99,Dhurandhar}. Much of the work to date has focused on time-delay reconstruction of the TDI variables from one-way measurements using six independently frequency-stabilized lasers. More recently, a configuration in which the lasers were phase-locked in a simple series to a stabilized master was analyzed \cite{TintoPRD03}. Although this analysis showed that time-delay interferometry is compatible with the phase-locking approach, it found no significant advantages of phase-locking compared to the one-way method.

This paper introduces a phase-locking configuration in which the phase of the light received from one interferometer arm is transferred to the phase of the light transmitted along an adjacent interferometer arm. This phase-locking arrangement allows two of the three LISA spacecraft to be considered as amplifying mirrors at a $30^{o}$ angle of incidence. This configuration is in contrast to the baseline \cite{LISA,TintoPRD03} phase-locking arrangement in which certain spacecraft appear as amplifying retro-reflecting mirrors. This Sagnac phase-locking configuration is discussed in section \ref{Phaselocking}.

Non-rotating Sagnac interferometers are insensitive to laser frequency noise because the counter-propagating beams have exactly matched optical paths. In section \ref{Sagnac} we note that orbital motion of LISA will result in a 14~km optical path length difference between the clockwise- and counterclockwise-propagating beams. Under these circumstances the TDI Sagnac readouts $\alpha$, $\beta$, $\gamma$ and $\zeta$ are no longer free from laser frequency noise. This arm length mismatch would lead to a significant degradation of the LISA sensitivity given expected levels of frequency stabilization. The implications for laser frequency stability are discussed, as is a post-processing algorithm that removes laser frequency noise. It is shown that unequal-arm post processing algorithms, similar to those used to obtain the Michelson variables $X$, $Y$, and $Z$ \cite{TintoPRD99}, can be applied to obtain frequency noise-free combinations of Sagnac-like variables.

\begin{figure}
\centerline {
\includegraphics[width=5.5in]{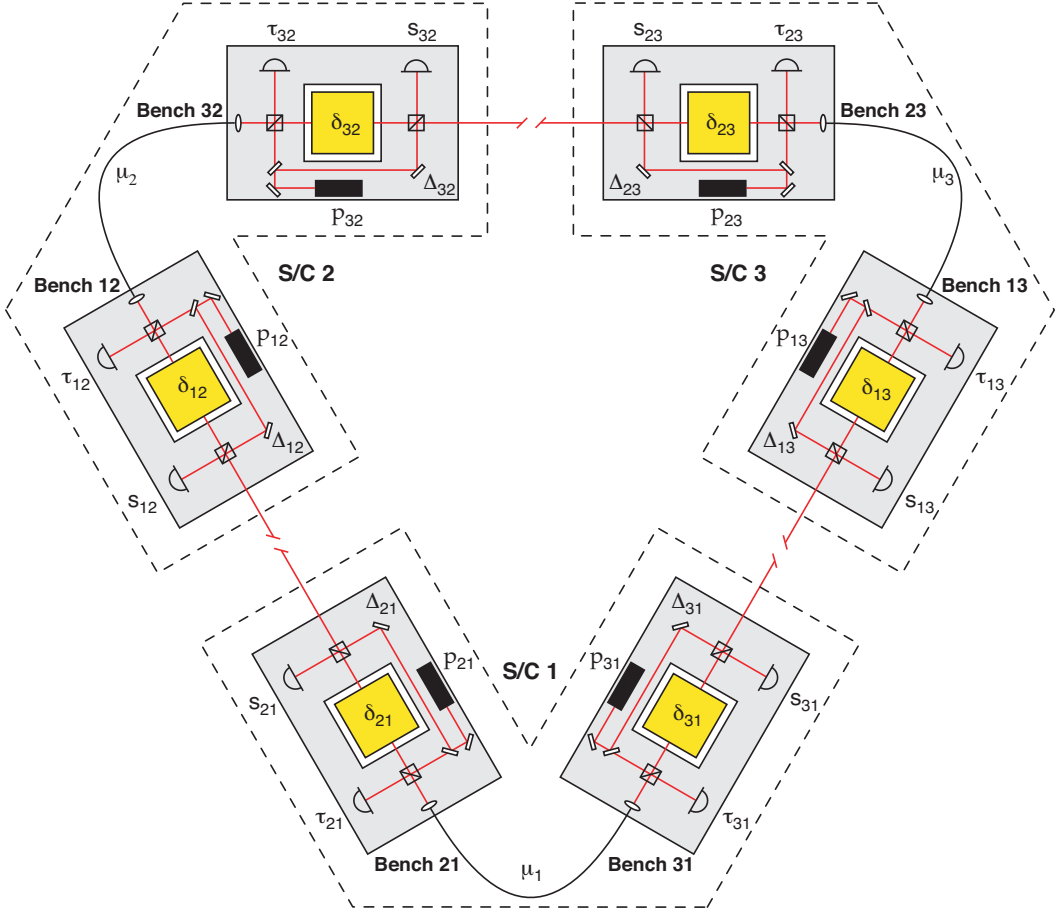}
}
\begin{center}{\caption{Simplified optical layout of the LISA interferometer. S/C: spacecraft, $s_{ij}$: inter-spacecraft phase measurement, $\tau_{ij}$: intra-spacecraft phase measurement, $p_{ij}$: laser phase, $\delta_{ij}$: proof mass displacement, $\Delta_{ij}$: optical bench displacement.}
\label{LISAbenches}}
\end{center}
\vspace{.2in}
\end{figure}

\section{Notation and phase measurements}
\label{nomenclature}

This paper uses a modified notation for describing the LISA parameters. The notation attempts to combine the most useful aspects of the three commonly used notations \cite{TintoPRD99,Dhurandhar,Hellings} in a consistent manner. Consider the diagram of the LISA interferometer in Fig. \ref{LISAbenches}, showing three spacecraft, S/C 1, S/C 2, and S/C 3, each containing two optical benches, two lasers and two proof masses. The propagation distance from S/C $i$ to S/C $j$ is denoted by $L_{ij}$ (with the displacement vector denoted by $\vec{L}_{ij}$). Since the spacecraft can move significantly over the light's  propagation time, the apparent spacecraft separation depends on the propagation direction ($L_{ij}\neq L_{ji}$). The LISA gravitational wave signal readouts are constructed from twelve phase measurements: six inter-spacecraft measurements taken at the front of the optical benches, denoted by $s_{ij}$, and six intra-spacecraft measurements taken at the back of the optical benches, denoted by $\tau_{ij}$. The first subscript of $s_{ij}$ is the distant S/C number from where the light originated and the second subscript is the local S/C number at which the measurement is made. The measurement $\tau_{ij}$ inherits the same subscript notation as the $s_{ij}$ measurement made on the same bench. We extend this subscript nomenclature to all other parameters on the bench including the laser phase, $p_{ij}$, the proof mass displacement, $\delta_{ij}$, the optical bench displacement, $\Delta_{ij}$ and the designator of the bench itself. Note that $\delta_{ij}$ and $\Delta_{ij}$ are scalar parameters representing the component of the displacement along the interferometer arm, defined to be positive for displacement in the direction of the distant spacecraft. Furthermore, $\Delta_{ij}$ and $\delta_{ij}$ are normalized to have units of phase.

The phase accumulated in traversing the optical fiber linking two benches on S/C $j$ is denoted by $\mu_{j}$. We assume the optical path through the fiber is independent of the propagation direction. This independence will be a requirement, not only for TDI, but for the operation of LISA in general.

The gravitational wave signal will modify the phase of the light traversing the interferometer arms. This phase shift is denoted by $h_{ij}$ for the light travelling from S/C $i$ to S/C $j$ where positive $h_{ij}$ represents a compression of the arm. Finally, the noise in the measurement of $s_{ij}$ is denoted by $n_{ij}$. This noise source has the same units as $s_{ij}$ and is typically a combination of shot noise and fluctuations in the optical path. Traditionally this noise source is referred to simply as optical path noise.

Following Dhurandhar \emph{et al.$\;$}\cite{Dhurandhar}, we adopt a delay operator, $D_{ij}$, to represent a time-delayed variable or parameter. The delay operator is defined by, 
\begin{eqnarray}
D_{ij}a(t)&\equiv& a(t-L_{ij}/c).
\end{eqnarray}
where $a(t)$ is an arbitrary function of time. This delay operator represents either a physical delay (due to propagation along an arm) or a post-processing delay (when data is stored in memory). Delay operators commute with each other and can be readily factorized to simplify data combinations \cite{Dhurandhar}.

The expressions for the six inter-spacecraft phase measurements can be obtained from examination of Fig. \ref{LISAbenches}.
\begin{eqnarray}
s_{21}&=&D_{21}p_{12}-p_{21}+D_{21}\Delta_{12}-\Delta_{21}+2\delta_{21}+h_{21}+n_{21}\label{s21} \\
s_{31}&=&D_{31}p_{13}-p_{31}+D_{31}\Delta_{13}-\Delta_{31}+2\delta_{31}+h_{31}+n_{31} \\
s_{12}&=&D_{12}p_{21}-p_{12}+D_{12}\Delta_{21}-\Delta_{12}+2\delta_{12}+h_{12}+n_{12} \\
s_{32}&=&D_{32}p_{23}-p_{32}+D_{32}\Delta_{23}-\Delta_{32}+2\delta_{32}+h_{32}+n_{32} \\
s_{23}&=&D_{23}p_{32}-p_{23}+D_{23}\Delta_{32}-\Delta_{23}+2\delta_{23}+h_{23}+n_{23} \\
s_{13}&=&D_{13}p_{31}-p_{13}+D_{13}\Delta_{31}-\Delta_{13}+2\delta_{13}+h_{13}+n_{13}\label{s13}
\end{eqnarray}
The measurements at the back of the proof masses are always taken as differences in order to cancel the otherwise overwhelming fiber noise.
\begin{eqnarray}
\tau_{31}-\tau_{21}&=&2(p_{21}-p_{31}-\delta_{21}+\delta_{31}+\Delta_{21}-\Delta_{31})\\
\tau_{12}-\tau_{32}&=&2(p_{32}-p_{12}-\delta_{32}+\delta_{12}+\Delta_{32}-\Delta_{12})\\
\tau_{23}-\tau_{13}&=&2(p_{13}-p_{23}-\delta_{13}+\delta_{23}+\Delta_{13}-\Delta_{23})
\end{eqnarray}
Note that there is no optical path noise contribution to the $\tau_{ij}$ measurements. Shot noise can be neglected, as ample power is present in both beams, and phase changes due to motion of the elements on the optical bench can be included in the $n_{ij}$ noise terms in equations \ref{s21}-\ref{s13} without loss of generality. For simplicity, the expressions above do not include clock noise. Clock noise unavoidably enters during the phase measurement process by an amount proportional to the beat-note frequency \cite{TintoPRD02}.

\section{Sagnac interferometer with phase-locked lasers}
\label{Phaselocking}
This section presents a phase-locking configuration that closely emulates a standard Sagnac interferometer. The phase-locking ensures that the phase of the light received at one bench is transferred to the phase of the light transmitted from the adjacent bench.

One of the lasers, laser 31 say, is chosen as the master laser and is frequency stabilized to an optical cavity (or other suitable frequency reference). The following phase-locking conditions are imposed by feeding back to the phases of the remaining five lasers.
\begin{eqnarray}
\tau_{31}-\tau_{21}&=&0\;\;\;\mbox{by feedback to }p_{21}\label{lock1}\\
s_{12}-\frac{\tau_{12}-\tau_{32}}{2}&=&0\;\;\;\mbox{by feedback to }p_{32}\label{lock2} \\
s_{23}-\frac{\tau_{23}-\tau_{13}}{2}&=&0\;\;\;\mbox{by feedback to }p_{13}\label{lock3}\\
s_{13}+\frac{\tau_{23}-\tau_{13}}{2}&=&0\;\;\;\mbox{by feedback to }p_{23}\label{lock4}\\
s_{32}+\frac{\tau_{12}-\tau_{32}}{2}&=&0\;\;\;\mbox{by feedback to }p_{12}\label{lock5}
\end{eqnarray}
The first locking condition (equation \ref{lock1}) phase-locks laser 21 to the master laser, ensuring their phases are related by,
\begin{eqnarray}
p_{21}=p_{31}+\delta_{21}-\delta_{31}-\Delta_{21}+\Delta_{31}.
\end{eqnarray}
Following the clockwise propagating beam from S/C 1 to S/C 2, equation \ref{lock2} implies the following condition for the phase of laser 32,
\begin{eqnarray}
p_{32}&=&D_{12}p_{21}+D_{12}\Delta_{21}-\Delta_{32}+\delta_{12}+\delta_{32}+h_{12}+n_{12}. \label{p32lock}
\end{eqnarray}
Equation \ref{p32lock} implies that the locking transfers the phase information of light received from S/C 1,  $D_{12}p_{21}+D_{12}\Delta_{21}+h_{12}$, to the light transmitted towards S/C 3, $p_{32}+\Delta_{32}$, with an error equal to the measurement noise, $\epsilon$. 
\begin{eqnarray}
\epsilon&=&\delta_{12}+\delta_{32}+n_{12} \label{mnoise}
\end{eqnarray}
Forcing the phase difference between laser 12 and the received light to equal the phase difference between laser 12 and the transmitted light, automatically ensures the phases of the received and transmitted light are equal (to within the measurement noise). Note that this phase-locking is independent of the phase and frequency of laser 12. 

Continuing to follow the clockwise propagating beam, the phase-locking condition of equation \ref{lock3} is imposed at S/C 3. This condition transfers the phase of the light received from S/C 2 to the phase of the light transmitted to S/C 1 (again to within the measurement noise). The phase of the light arriving at S/C 1 relative to the phase of the local laser, $s_{31}$, is measured.

The remaining two locking conditions, equations \ref{lock4} and \ref{lock5}, preserve the phase information for the counterclockwise propagating beam. With all locking conditions satisfied, S/C 2 and S/C 3 behave as amplifying mirrors at an angle of incidence of $30^{o}$. In this situation, the LISA constellation can be treated as a simple Sagnac interferometer and the phase difference between the clockwise and counterclockwise propagating beams, $\alpha$, is simply,
\begin{eqnarray}
\alpha&=&s_{31}- s_{21}\label{alphasimp}
\end{eqnarray}

This expression is equivalent to the TDI Sagnac variable $\alpha$ formed from one-way measurements \cite{ArmstrongAPJ99} shown in equation \ref{onewayalpha}. The five locking conditions ensure that the quantities inside the square brackets vanish. 
\begin{eqnarray}
\alpha&=&\;\;s_{31}+ D_{31}\left[s_{23} -\frac{\tau_{23}-\tau_{13}}{2}\right] + D_{23}D_{31}\left[s_{12} - \frac{\tau_{12}-\tau_{32}}{2}\right]  \nonumber \\
&&- s_{21}- D_{21}\left[s_{32} + \frac{\tau_{12}-\tau_{32}}{2}\right] - D_{32}D_{21}\left[s_{13} + \frac{\tau_{23}-\tau_{13}}{2}\right] \nonumber \\
&&- (1 + D_{12}D_{23}D_{31}) \left[\frac{\tau_{31}-\tau_{21}}{2}\right] \label{onewayalpha}
\end{eqnarray}

If the interferometer arm lengths are independent of the propagation direction, i.e. $L_{ij}=L_{ji}$, then the laser frequency noise and bench noise are cancelled and only the proof mass noise, optical path noise and gravitational wave signals remain.
\begin{eqnarray}
\alpha&=&(1 - D_{12}D_{31} D_{23}) (\delta_{31}-\delta_{21})+(D_{31} - D_{12} D_{23}) (\delta_{13}+\delta_{23})-(D_{12} - D_{31} D_{23}) (\delta_{12}+\delta_{32}) \nonumber \\
 &&+(1- D_{12} D_{23}) h_{31}-(1-D_{31} D_{23}) h_{12} + (D_{31}  - D_{12}) h_{23}\nonumber \\
&& +D_{31} D_{23} n_{12} - D_{12} D_{23} n_{13} - n_{21} + D_{31} n_{23} + n_{31} - D_{12} n_{32}
\end{eqnarray}

Using this phase-locking approach to obtain $\alpha$ has several clear advantages. The measurements are taken simultaneously, thus no time delays are required and there is no need for knowledge of the absolute lengths of the arms. Both measurements are taken at S/C 1 and so neither clock synchronization nor exchange of phase measurements between spacecraft is necessary. Although not included in the calculation above, it can be shown that $\alpha$ is also immune to clock noise. Clock noise enters into a phase measurement by an amount proportional to the beat-note frequency. If the LISA constellation has a constant angular velocity then the doppler shifts of the clockwise and counterclockwise beams are identical. Therefore the clock noise in $s_{21}$ and $s_{31}$ will be equal and will cancel when the signals are subtracted to give $\alpha$. The S/C 2 and S/C 3 clock noise also cancels because the phase-locking ensures the beat-note frequencies are the same on all four photodetectors.

The advantages mentioned above apply only to the Sagnac variable centered on the master laser's spacecraft (e.g. $\alpha$ if the master is on S/C 1). The other TDI Sagnac readouts can still be reconstructed, however, with slightly simpler expressions than if the one-way method were used. For example, the expression for $\beta$ in terms of one-way measurements is,
\begin{eqnarray}
\beta&=&s_{12}+ D_{12} \left[s_{31} - \frac{\tau_{31}-\tau_{21}}{2}\right] + D_{12} D_{31} \left[s_{23} -  \frac{\tau_{23}-\tau_{13}}{2}\right]  \nonumber \\ 
&& - s_{32} - D_{23} \left[s_{13} + \frac{\tau_{23}-\tau_{13}}{2}\right] - D_{31} D_{23} \left[ s_{21} +\frac{\tau_{31}-\tau_{21}}{2}\right]  \nonumber \\
&& - (1 + D_{12} D_{23} D_{31})\frac{\tau_{12}-\tau_{32}}{2} \label{beta}
\end{eqnarray}
A similar expression is obtained for $\gamma$ by cyclic permutation of the subscripts $1\rightarrow2\rightarrow3\rightarrow1$ \cite{TintoPRD99}. With the phase-locking implemented equation \ref{beta} simplifies to,
\begin{eqnarray}
\beta&=&D_{12}s_{31} - D_{31} D_{23}  s_{21} + (1- D_{12} D_{23} D_{31})s_{12}.
\end{eqnarray}
Time shifting of data and knowledge of the arm length are now needed to reconstruct $\beta$ and $\gamma$. The variables also utilize measurements made at two different spacecraft and so synchronization of the clocks and some exchange of phase information between spacecraft is needed. 

\section{Sagnac effect and the LISA constellation}
\label{Sagnac}
Previous work on time-delay interferometry variables has ignored the rotation of the LISA constellation. This section discusses the implications for the Sagnac variables if the rotation of the LISA constellation is included.

Sagnac interferometers are commonly used as rotation sensors. The difference between the optical paths of the counter-propagating beams in a rotating Sagnac interferometer is,
\begin{eqnarray}
\Delta L&=&L_{12}+L_{23}+L_{31}-(L_{13}+L_{32}+L_{21})\nonumber \\
&=&\frac{4\vec{\Omega} \cdot \vec{A}}{c}
\end{eqnarray}
where $\vec{\Omega}$ is the angular velocity, $\vec{A}$ is the area enclosed by the optical paths, and $c$ is speed of light. The dot product indicates that only the component of the angular velocity vector out of the plane contributes to the Sagnac effect. An angular velocity of 1~cycle/year is $2\times 10^{-7}$~rad/s. The LISA constellation will precess in the plane of the interferometer with one cycle per year. The heliocentric orbital motion of the constellation must also be considered, as the plane of the interferometer is only $60^{o}$ from the ecliptic. The component of the angular velocity parallel to the area vector is therefore $[1-\cos(60^{o})]\cdot|\vec{\Omega}|=0.5$~cycles/year~$=10^{-7}$~rad/s. 

An equilateral triangle with sides of length $L$ has an area given by,
\begin{eqnarray}
|\vec{A}|=\frac{\sqrt{3}L^{2}}{4}.
\end{eqnarray}
Modeling LISA as an equilateral triangle of arm length $5\times 10^{9}$~m, the interferometer encloses an area of $1.1\times10^{19}$~m$^{2}$. With these assumptions $\Delta L=14.4$~km. An independent calculation of $\Delta L$ can be made using the initial positions and velocities of the spacecraft. Based on the initial conditions published by Folkner \cite{FolknerCQG2001}, a difference between the round trip optical paths of the counter-propagating beams of 14.1 km is expected during the first round-trip ($\sim$50 seconds). This calculation assumed constant spacecraft velocities and that changes in the spacecrafts' positions over 50 seconds were negligible compared to the spacecraft separation.

\begin{figure}[htb]
\centerline {
\includegraphics[width=4in]{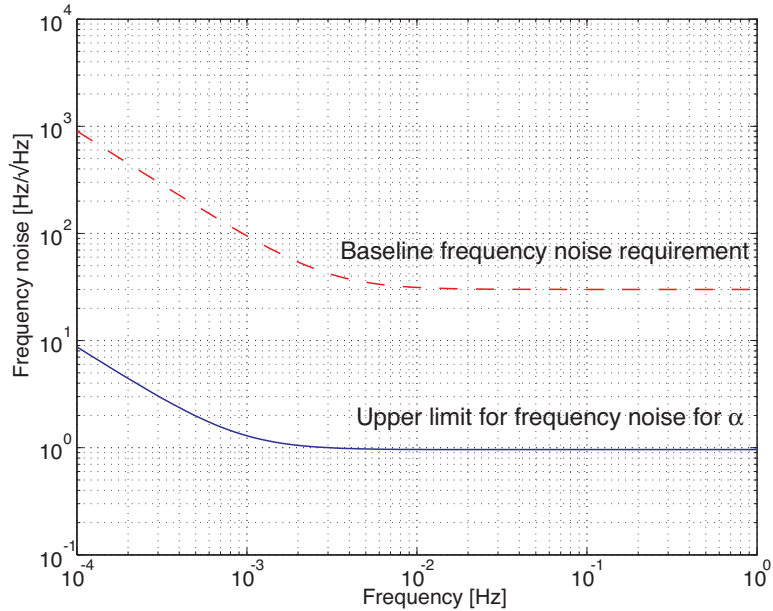}
}
\begin{center}{\caption{Laser frequency noise requirements for LISA. Dashed line: baseline frequency noise requirement of $\sqrt{1+(3{\rm ~mHz}/f)^{2}}\times30$~Hz/$\sqrt{\rm Hz}$ \cite{Guido}. Solid line: laser frequency noise needed to equal the combined proof mass and optical path noise in (the unequal-arm) $\alpha$.}
\label{freqnoisereq}}
\end{center}
\end{figure}

In previous calculations, the arm length difference introduced by the Sagnac effect was not considered and it was assumed that $L_{ij}=L_{ji}$. Given that the $\alpha$, $\beta$ and $\gamma$ variables are simply time-delay reconstructed Sagnac interferometers it is reasonable to expect that they will be affected by this arm length difference. We will now re-examine these Sagnac variables for the general case, $L_{ij}\neq L_{ji}$.
\begin{eqnarray}
\alpha&=&(D_{12} D_{23} D_{31} - D_{13} D_{21} D_{32}) (p_{31}+\Delta_{31})\nonumber \\
&&
-(D_{21} - D_{23} D_{31}) (\delta_{12}+ \delta_{32})
+ (D_{31} - D_{21} D_{32}) (\delta_{13}+\delta_{23})
+ (1 - D_{12}D_{23} D_{31}) (\delta_{31}- \delta_{21}) \nonumber \\ &&
+ (h_{31}+n_{31}) + D_{31} (h_{23}+n_{23})+ D_{23} D_{31} (h_{12}+n_{12})\nonumber \\ &&
 - (h_{21}+n_{21})   - D_{21}( h_{32}+n_{32})- D_{21} D_{32}( h_{13}+n_{13})\label{alpha_unequal}
\end{eqnarray}
The first line of equation \ref{alpha_unequal} shows that the laser phase noise and optical bench noise \cite{EstabrookPRD00} no longer cancel, coupling into the Sagnac output proportionately to the difference in delays. In the frequency domain, these noises enter the Sagnac output with a magnitude of $|\sin(2\pi f\Delta L/c)|$, where $\Delta L=1.4\times10^{4}$~m.  This level of sensitivity to the bench noise is probably acceptable for frequencies below 1~Hz. For example, bench noise with a white spectral density of 10~nm/$\sqrt{\rm Hz}$ would contribute 3~pm/$\sqrt{\rm Hz}$ of displacement noise at 1~Hz and only 3~fm/$\sqrt{\rm Hz}$ at 1~mHz. However, the presence of laser frequency noise is of greater concern. Figure \ref{freqnoisereq} shows a comparison of the current laser frequency noise requirement (upper line) and the frequency noise needed to equal the combined budget for proof mass and optical path noise (lower line). To produce the lower line it was assumed that each proof mass contributes 10~pm$\times (3{\rm ~mHz}/f)^{2}/\sqrt{\rm Hz}$ and each of the six measurements is subject to an optical path noise of 20~pm/$\sqrt{\rm Hz}$. Note that this is the frequency stability required to equal the dominant noise sources. In practice, the frequency noise requirement will need to be more stringent to avoid degradation of the LISA sensitivity and the ``upper limit'' of Fig. \ref{freqnoisereq} would be replaced by a requirement level lower by a factor of 2, approximately. Augmenting the laser frequency stabilization system with a second loop using the 5 million kilometer arms as a reference is a promising approach for achieving better stability.

An alternative solution to improving the frequency stability is to modify the TDI combinations to obtain frequency noise-free variables in the presence of rotation. Although time-delay interferometry was originally conceived to remove laser frequency noise from an unequal-arm Michelson interferometer \cite{TintoPRD99} the same procedure can be applied to an unequal-arm Sagnac interferometer. We define a new set of modified Sagnac variables, $\alpha_{1}$, $\alpha_{2}$, and $\alpha_{3}$, representing Sagnac-like TDI variables beginning and ending at S/C 1, 2, and 3 respectively.  The variable $\alpha_{1}$ is obtained from $\alpha$ by delaying the clockwise measurements by the counterclockwise roundtrip time, and delaying the counterclockwise measurements by the clockwise roundtrip time, before combining them with the original $\alpha$ variable.
\begin{eqnarray}
\alpha_{1}&=&(1 - D_{13 }D_{21} D_{32})\left(s_{31}+D_{31} \left[s_{23}-\frac{\tau_{23}-\tau_{13}}{2}\right]+D_{23} D_{31} \left[s_{12}-\frac{\tau_{12}-\tau_{32}}{2}\right]\right) \nonumber \\
&&-  (1 - D_{12 }D_{23} D_{31}) \left(s_{21}+ D_{21}\left[s_{32 }+\frac{\tau_{12}-\tau_{32}}{2}\right]+D_{21}D_{32} \left[s_{13}+\frac{\tau_{23}-\tau_{13}}{2}\right]\right) \nonumber \\
&&- (1 - D_{12 }D_{13} D_{21} D_{23} D_{31} D_{32}) \left[\frac{\tau_{31}-\tau_{21}}{2}\right] \label{alpha1}
\end{eqnarray}

If the phase-locking of section \ref{Phaselocking} is implemented, the quantities inside the square brackets vanish and equation \ref{alpha1} simplifies to,
\begin{eqnarray}
\alpha_{1}&=&(1 - D_{13 }D_{21} D_{32})s_{31}-  (1 - D_{12 }D_{23} D_{31})s_{21} \label{alpha1simp}
\end{eqnarray}

The symmetrized Sagnac, $\zeta$, has the same problem in that it is also sensitive to laser frequency noise and bench noise.  This was first calculated by Hellings \cite{Hellings2003}. However, Tinto \emph{et al.} \cite{TintoPRD03b} pointed out that $\zeta$ can be modified to produce three combinations in which the frequency noise and bench noise are cancelled. In contrast to the symmetric variable $\zeta$, these three combinations, denoted by $\zeta_{1}$, $\zeta_{2}$, and $\zeta_{3}$, combine the data from the three spacecraft asymmetrically. The expression for $\zeta_{1}$ is presented below with $\zeta_{2}$, and $\zeta_{3}$ obtained by cyclic permutation of the subscripts.
\begin{eqnarray}
\zeta_{1}&=&(D_{13} D_{21} -D_{23}) (D_{31} s_{12}+D_{32}s_{31}-D_{31}s_{32}) 
- (D_{12}D_{31} - D_{32})(D_{21}  s_{13}+D_{23} s_{21}-D_{21} s_{23} )
\nonumber \\
&&- (D_{13} D_{21} D_{31} - D_{23} D_{31} +D_{12} D_{21} D_{23} D_{31} - D_{21} D_{23} D_{32}) \frac{\tau_{12}-\tau_{32}}{2}\nonumber \\ 
&&- (D_{12} D_{21} D_{31} - D_{21} D_{32} +D_{13} D_{21} D_{31} D_{32} - D_{23} D_{31} D_{32}) \frac{\tau_{23}-\tau_{13}}{2}\nonumber \\
&& - (D_{12} D_{13} D_{21} D_{31} - D_{23} D_{32})\frac{\tau_{12}-\tau_{32}}{2} 
\end{eqnarray}

Given the extra complexity in constructing the frequency noise-free Sagnac variables in the presence of rotation, the Michelson variables now seem more appealing. The TDI Michelson parameters are immune to the Sagnac effect as the interferometers contain zero area. The expressions for $X$, $Y$ and $Z$ remain unchanged apart from allowing for different delays of the counter-propagating beams. These expressions, denoted by $X_{1}$, $X_{2}$, and $X_{3}$ for Michelson-type readouts centered on S/C 1, 2, and 3 respectively, have been independently derived by Hellings \cite{Hellings2003} and Tinto \emph{et al.$\;$}\cite{TintoPRD03b}. These TDI Michelson readouts have a slightly better gravitational wave sensitivity than the Sagnac outputs particularly around $3$~mHz. The reason for this sensitivity difference is that in a Sagnac interferometer both the gravitational wave signal and proof mass noise are suppressed below 20~mHz, whereas the optical path noise is not. In the TDI Michelson outputs the gravitational wave signal, proof mass noise and optical path noise are suppressed equally at low frequencies.  One minor drawback of the TDI Michelson readouts is a slight increase in the noise due to currently envisaged clock noise correction schemes. The extra shot noise added by the clock noise correction algorithm was ignored in previous calculations and reduces the sensitivity to gravitational waves at frequencies around the 30~mHz and its harmonics \cite{Shaddock}. Some amount of extra noise will also be added to $\alpha_{1}$, $\alpha_{2}$ and $\alpha_{3}$ if clock noise correction is used.

One of the most appealing aspects of the Sagnac approach is that the symmetrized Sagnac, $\zeta$, provided a measurement of the instrumental noise floor with a much reduced sensitivity to gravitational waves. This is potentially a valuable tool for differentiating the gravitational wave background from the instrumental noise. Another variable with significantly reduced sensitivity to the gravitational wave signal could be constructed by summing the three TDI Michelson outputs.
\begin{eqnarray}
\Sigma&=&X_{1}+X_{2}+X_{3} \label{sigma}
\end{eqnarray}
In the long wavelength limit, $\Sigma$ is insensitive to gravitational waves and is the TDI Michelson analog of the Sagnac-like variable $T$ \cite{PrincePRD02}. The summation in equation \ref{sigma} could be performed on Earth with relatively relaxed timing requirements as the laser frequency noise is not present in the TDI Michelson variables $X_{1}$, $X_{2}$, and $X_{3}$.

\section{Summary}

A phase-locking configuration has been identified that offers several potential advantages 
for the Sagnac  mode of operation of LISA. For a static LISA constellation, this configuration provides one Sagnac output that is free from laser frequency noise, bench noise, and clock noise. Furthermore, no clock synchronization, arm length knowledge, time shifting of data or exchange of phase information between spacecraft is required. However, we have demonstrated that the orbital motion of the LISA spacecraft will break the optical path-length symmetry between counter-propagating beams. This places new constraints on the laser frequency noise for the baseline Sagnac configuration. Alternatively, new time-delayed combinations of the phase measurements can be introduced that provide laser frequency noise-free outputs for LISA.

\section*{Acknowledgments}
The author thanks Robert Spero, Andreas Kuhnert and Charles Harb for many useful discussions. The author has also benefited from discussions with Massimo Tinto, John Armstrong, and Frank Estabrook, who derived the expressions for $\zeta_{1}$, as well as discussions with Ron Hellings who first calculated that $\zeta$ is also sensitive to laser frequency noise if rotation is considered. This research was performed at the Jet Propulsion Laboratory, California Institute of Technology, under contract with the National Aeronautics and Space Administration.

 \end{document}